\documentclass{elsart}

\usepackage{graphicx}
\usepackage{amssymb}

\begin{document}
\begin{frontmatter}
\title{Fast convergence of path integrals for many-body systems}
\author{A. Bogojevi\' c\corauthref{alex}},
\ead{alex@phy.bg.ac.yu} \corauth[alex]{Corresponding author.}
\author{I. Vidanovi\' c},
\author{A. Bala\v z}, \and
\author{A. Beli\'c}
\address{Scientific Computing Laboratory, Institute of Physics Belgrade\\
Pregrevica 118, 11080 Belgrade, Serbia\\
http://scl.phy.bg.ac.yu/}

\begin{abstract}
We generalize a recently developed method for accelerated Monte Carlo calculation of path integrals to the physically relevant case of generic many-body systems. This is done by developing an analytic procedure for constructing a hierarchy of effective actions leading to improvements in convergence of $N$-fold discretized many-body path integral expressions from $1/N$ to $1/N^p$ for generic $p$. In this paper we present explicit solutions within this hierarchy up to level $p=5$. Using this we calculate the low lying energy levels of a two particle model with quartic interactions for several values of coupling and demonstrate agreement with analytical results governing the increase in efficiency of the new method. The applicability of the developed scheme is further extended to the calculation of energy expectation values through the construction of associated energy estimators exhibiting the same speedup in convergence.
\end{abstract}

\end{frontmatter}

\section{Introduction}

Originally introduced in quantum mechanics \cite{feynman,feynmanhibbs} and later most widely used in high energy theory \cite{itzyksonzuber,coleman} and condensed matter physics \cite{parisi,itzyksondrouffe}, path integrals have become important tools throughout the physical sciences from atomic, molecular and nuclear physics, to chemistry and biophysics. Moreover, path integrals are starting to play important roles in several areas of mathematics and in modern finance \cite{baaquie}, especially in option pricing applications \cite{kleinert1,kleinert2}. The downside of the formalism is that the mathematical properties of path integrals are not sufficiently well understood, and that an extremely small number of path integrals can be solved exactly \cite{steinergrosche}. An extensive overview of the path integral formalism and its various applications can be found in \cite{kleinertbook}.

Beside their key position in analytical approaches to quantum theory, path integrals have an important role in direct numerical simulations of realistic many-body systems. Such simulations have made possible practical comparisons of the predictions of various theoretical models with the results of associated experiments. Further, they have led to a deeper understanding of the complex physical phenomena involved \cite{ceperley}.

The starting point in all such calculations is the time-sliced expression for the general quantum-mechanical transition amplitude \cite{feynmanhibbs},
\begin{equation}
A_N(a,b;T)=\frac{1}{(2 \pi \varepsilon)^{\frac{M N d}{2}}}\int d q_1\cdots d q_{N-1}\ e^{-S_N}.\
\label{eq:base}
\end{equation}
from initial state $|a\rangle$ to final state  $|b\rangle$, for time interval $T$, where $N$ is the number of time slices, $\varepsilon=T/N$ and $S_N$ is the naively discretized action for a system of $M$ non-relativistic particles in $d$ spatial dimensions. The $N\rightarrow \infty$ limit of the above discretized amplitude gives the continuum amplitude $A(a,b;T)=\langle b|e^{-\hat H T}|a\rangle$. The evaluation of these types of discretized expressions is handled by a variety of well developed numerical integration methods, and this is one of the principle reasons for the popularity of the path integral approach in numerical simulations.

The performance of numerical algorithms for the calculation of path integrals is determined by the efficiency of the applied integration techniques, as well as by the speed of convergence of the discretized amplitudes to the continuum. The first aspect has been successfully dealt with by numerous Monte Carlo integration methods based on efficient sampling of trajectories. The second is usually addressed by the use of better discretizations, i.e by substituting better effective actions for the naively discetized action. The underlying idea here is to construct and use effective actions that lead to improved convergence of physical expressions to the (same) continuum limit.

Typically, physical expression calculated using the naively discretized action converge to the continuum as $1/N$. A review of a variety of effective actions constructed by improving short-time propagation and using generalizations of the Trotter-Suzuki formula \cite{suzuki}, together with their ranges of application and dependence on ordering prescriptions, are given in \cite{bogojevicprb}. For a long time now, the state of the art result has been the $1/N^4$ convergence of discretized partition functions obtained by Takahashi and Imada \cite{takahashiimada}, and Li and Broughton \cite{libroughton}. In these papers the authors used a generalized form \cite{deraedt2} of the Trotter formula and the cyclic property of the trace to increase the speed of convergence of the discretized partition functions to $1/N^4$. We stress that this increase in the speed of convergence only holds for partition functions and not for amplitudes. Some recent results regarding the efficient implementation of PIMC algorithms may be found in refs. \cite{predescu1,predescu2,predescu3,yamamoto,paulinetall}. In general, the main feature of the improved effective action -- faster approach to the continuum limit, i.e with smaller number of time slices -- translates into numerical speedup, simpler integration and smaller statistical fluctuations of the quantities calculated.

In a recent paper \cite{bogojevicpre}, we have introduced the concept of the ideal effective action $S_N^*$. Discretized amplitudes calculated with the ideal effective action do not depend on the discretization coarseness $N$, i.e. they are all equal to the sought-after continuum amplitude
\begin{equation}
A_N^*(a,b;T)=A(a,b;T)\, .
\end{equation}
Note that for free particles $S_N^*$ is in fact just the naively discretized action. In \cite{bogojevicpla1} we derived the integral equation for the ideal action for a single particle moving in one spatial dimension in a general potential. We then solved this integral equation as an asymptotic expansion in $\varepsilon$. The obtained truncation of the ideal effective action $S_N^{(p)}$ to order $O(\varepsilon^p)$ systematically improves convergence to the continuum to $1/N^p$:
\begin{equation}
A_N^{(p)}(a,b;T)=A(a,b;T)+O(1/N^p)\, ,
\label{eq:impcon}
\end{equation}
where $p=1$ corresponds to the naive action. Up to now the work has focused on one particle theories in one dimension. Within this set of theories the outlined procedure has been used to determine explicit expressions for the set of effective discretized actions up to level $p=12$. The ensuing speedup of several orders of magnitude has been verified by extensive numerical simulations \cite{bogojevicprl}. The speedup holds for the calculation of all  path integrals -- for transition amplitudes, partition functions, energy expectation values (in combination with appropriate energy estimators), as well as for calculations of energy levels. In this paper we extend the above method to the physically relevant case of general many-particle non-relativistic systems in arbitrary number of spatial dimensions. This is done within a new analytical approach to the construction of the effective actions hierarchy. The new approach gives the same effective actions as the the old one when applied to one particle one dimensional theory, however its computational complexity is lower, making it possible to determine effective actions at higher $p$ levels. In particular, the new approach allows for a relatively straight-forward extension of the formalism to many-body theories and arbitrary dimensions.

The present paper gives the analytical derivation of these results, as well as a series of Monte Carlo simulations implementing the newly derived effective actions and explicitly displaying that the derived speedup indeed holds. The paper is organized as follows: Section \ref{sec:eff} introduces the hierarchy of effective discretized actions and gives the new and extended (many particles, higher dimensions) analytical approach to the construction of these effective actions. Section \ref{sec:num} presents numerical results that demonstrate the validity of the analytically derived speedup in convergence. The efficiency of the new approach is further demonstrated in Section \ref{sec:spec} through the calculation of energy spectra. A construction of virial energy estimators corresponding to newly derived effective actions and the results of numerical simulations that implement them are given in Section \ref{sec:est}.

We conclude the paper with a brief summary of obtained results, indicating what we see to be the next steps in our line of research and its future applications. Appendix A gives a list of integration formulas needed for calculations up to level $p=5$. Appendix B lists the corresponding effective actions representing the new state-of-the-art for path integral calculations of non-relativistic many-body systems in arbitrary number of dimensions. Higher level effective actions can be found on our web site \cite{speedup}.

\section{New approach to the derivation of effective actions}
\label{sec:eff}

We present a method for systematically increasing the efficiency of PIMC calculations of a non-relativistic quantum system consisting of $M$ distinguishable particles in $d$ spatial dimensions, with a Hamiltonian of the form $\hat H=\hat K+\hat V$, where $\hat K$ is the usual kinetic energy operator, and $\hat V$ is the potential describing inter-particle interactions and interactions with external fields. The above set of theories encompasses a large number of physically relevant models. However, as we shall see from the derivations, the method is in principle applicable to all quantum theories. For the considered class of theories, the naively discretized action  $S_N$ is
\begin{equation}
S_N=\sum_{n=0}^{N-1} \varepsilon \left(\sum_{\ell=1}^M \frac{1}{2}{\left(\frac{\delta_{n,\ell}}{\varepsilon}\right)}^2+ V(\bar{q}_n)\right)\, .
\label{eq:naive}
\end{equation}
Vectors with one index (e.g. $q_n$) represent the set of positions of all particles after $n$ time steps of length $\varepsilon=T/N$, while vectors with two indices (e.g. $q_{n,\ell}$) represent $d$-dimensional positions of particle $\ell$ at time step $n\varepsilon$. We have also introduced the associated discretized velocities $\delta_{n, \ell}=q_{n+1, \ell}-q_{n, \ell}$ and mid-point coordinates $\bar{q}_n=(q_n+q_{n+1})/2$. To summarize, $n$ counts the time steps, $\ell$ the different particles. In the most compact notation, we may think of a configuration (trajectory) of the quantum system as a single vector $q$ whose individual components $q_i$ take on $Md$ possible values, representing positions of all the particles at a given time step. In this notation the $N\to\infty$ limit of the above discretized amplitude is symbolically written as the path integral
\begin{equation}
A(a,b;T)=\int_{ q(0)= a}^{q(T)= b} [d q]\ e^{-S[q(t)]}\, .
\end{equation}
Note that we are using the mid-point ordering prescription and units in which $\hbar$ and the particle masses have been set to unity.

The defining relation for path integrals as the continuum limit of discretized amplitudes given by equation (\ref{eq:base}) follows from the completeness relation (decomposition of unity)
\begin{equation}
A(a,b;T)=\int d q_1\cdots d q_{N-1}\ A(a,q_1;\varepsilon)\cdots A(q_{N-1},b;\varepsilon)\, ,
\label{eq:com}
\end{equation}
through the substitution of short-time amplitudes $A(q_n,q_{n+1};\varepsilon)$ calculated to first order in time step $\varepsilon$, leading to the naive action in equation (\ref{eq:base}). This is what gives the $1/N$ convergence to standard path integral expressions. A faster converging result may be obtained by evaluating the amplitudes under the integral to higher orders in $\varepsilon$. From the above completeness relation, it follows that the ideal discretized action $S_N^*$ leads to exact propagation in time, and is given in terms of the exact amplitude, according to
\begin{equation}
A(q_n,q_{n+1};\varepsilon)=\left(2 \pi \varepsilon \right)^{-\frac{Md}{2}}\ e^{-S_n^*}\, .
\label{eq:ideal}
\end{equation}
The ideal discretized action $S_N^*$ is simply the sum of expressions $S_n^*$:
\begin{equation}
S_N^*=\sum_{n=0}^{N-1} S_n^{*}(q_n,q_{n+1};\varepsilon)\, .
\end{equation}

Full knowledge of the ideal action $S_N^*$ is equivalent to knowing the exact expression for all the amplitudes $A$. At first, this would seem to indicate that nothing new is to be gained by equation (\ref{eq:ideal}). This, however, is not the case. We will use equation (\ref{eq:ideal}) to input new analytical information into our numerical procedure by calculating amplitudes within some available analytical approximation scheme. We focus on the calculation of the short time propagation as a power series expansion in $\varepsilon$, which starts from the naive action (\ref{eq:naive}).
The details of this calculation have been inspired by a similar derivation given in \cite{kleinertbook}. One can also attack the problem of solving equation (\ref{eq:ideal}) using various other approximative schemes, especially in the case when short time approximation is not appropriate. The application of the Feynman and Kleinert variational approach \cite{feynmankleinert,kleinert3}, further developed in \cite{kleinert4,kleinertetall,weissbachetall,pelsteretall,brandtpelster}, could prove useful is such cases.

Improved $1/N^p$ convergence of path integrals follows from calculating the general short-time amplitude $A(q_n,q_{n+1};\varepsilon)$ up to the order $\varepsilon^p$. To this end, we first perform a simple shift of integration variable $q=\xi+x$ about a fixed referent trajectory $\xi$,
\begin{equation}
A(q_n,q_{n+1};\varepsilon)=e^{-S_n[\xi]} \int_{x(-\varepsilon/2)=0}^{x(\varepsilon/2)=0} [d x]\ e^{-\int_{-\varepsilon/2}^{\varepsilon/2}d s \left( \frac{1}{2} \dot{{x}}^2+U(x;\xi)\right)}\, .
\label{eq:1}
\end{equation}
We have also shifted the time from $t\in[n\varepsilon, (n+1)\varepsilon]$ to $s\in[-\varepsilon/2,\varepsilon/2]$. The referent trajectory $\xi$ satisfies the same boundary conditions as $q$. As a result, the new integration variable $x$ vanishes at the boundaries. The action $S_n[\xi]$ is defined as
\begin{equation}
S_n[\xi]=\int_{-\varepsilon/2}^{\varepsilon/2} d s \left(\frac{1}{2}\dot{\xi}^2+V(\xi)\right)\, ,
\end{equation}
and $U(x;{\xi})=V(\xi+x)-V(\xi)-x \ddot{\xi}$, with dots representing derivatives over time $s$. The amplitude may now be written as
\begin{equation}
A(q_n,q_{n+1};\varepsilon)=\frac{e^{-S_n[\xi]}}{(2 \pi \varepsilon)^{\frac{M d}{2}}}\ \left\langle e^{ -\int_{-\varepsilon/2}^{\varepsilon/2} d s\ U(x;\xi)}\right\rangle\, ,
\label{eq:main1}
\end{equation}
where $\langle ... \rangle$ denotes the expectation value with respect to the free particle action. The above expression holds for any choice of referent trajectory $\xi$. Equations (\ref{eq:ideal}) and (\ref{eq:main1}) now give the expression for the ideal effective action
\begin{equation}
S_n^*=S_n[\xi]-\ln\left\langle e^{-\int_{-\varepsilon/2}^{\varepsilon/2}ds\ U(x;\xi)}\right\rangle\, .
\label{eq:formal}
\end{equation}

The class of theories considered is free of ordering ambiguities, i.e. different ordering prescriptions yield the same continuum result. For concreteness we work in the mid-point prescription, calculating the above amplitude as a power series in time step $\varepsilon$. The free particle expectation value in equations (\ref{eq:main1}) and (\ref{eq:formal}) is calculated using a Taylor expansion in powers of $U$:
\begin{eqnarray}
\left\langle e^{ -\int d s\ U(x;\xi)}\right\rangle&=&1-\int d s\left\langle U(x;\xi)\right\rangle+\nonumber\\
&&\qquad +\frac{1}{2} \int \int d s d s' \left\langle U(x;\xi)U(x';\xi' )\right\rangle+\ldots
\end{eqnarray}
with shorthand notation $x'=x(s'), \xi'=\xi(s')$. By expanding $U(x;\xi)$ around the referent trajectory $\xi$, we get
\begin{equation}
U(x;\xi)=x_i \left(\partial_i V(\xi)-\ddot{\xi}_i\right)+\frac{1}{2} x_i x_j \partial_i \partial_j V(\xi)+\ldots
\label{eq:Uexp}
\end{equation}
From now on we assume summation over repeated indices. The expectation values of products $\langle x_i(s)\ldots x_j(s') \rangle$ are now calculated in the standard way with the help of the free-particle generating functional given in terms of the propagator:
\begin{eqnarray}
\Delta(s,s')_{ij} =&& \frac{\delta_{ij}}{\varepsilon}\,\theta(s-s')\left(\frac{\varepsilon}{2}-s\right)\left(\frac{\varepsilon}{2}+s'\right)\nonumber\\
&&\qquad +\frac{\delta_{ij}}{\varepsilon}\,\theta(s'-s)\left(\frac{\varepsilon}{2}+s\right)\left(\frac{\varepsilon}{2}-s'\right).
\label{eq:propagator}
\end{eqnarray}
From this follow the usual Wick's theorem results $\langle x_i(s)\rangle=0$, $\langle x_i(s) x_j(s')\rangle=\Delta(s,s')_{ij}$, etc. Note that the generating functional (and so all the expectation values) is independent of the specific choice of referent trajectory, i.e. the boundary conditions for the $x$ are the same for all choices of $\xi$, and so the propagator is always given by (\ref{eq:propagator}). Different choices of $\xi$ simplify  different approximation schemes: using the classical trajectory for $\xi$ is optimal for recovering semi-classical expansion, the choice of linear referent trajectories will turn out to be optimal for short-time expansion.

We next wish to perform the remaining integrations over $s$. Because of the explicit dependence of the referent trajectory on $s$, we first expand the potential and all its derivatives in (\ref{eq:Uexp}) around some reference point. The choice of $\bar{{q}}_n$ for that point corresponds to the mid-point ordering prescription. Once one chooses the referent trajectory $\xi(s)$, all expectation values in (\ref{eq:main1}) are given in terms of quadratures. The choice of linear referent trajectories $\xi(s)=\bar{q}_n+\frac{\delta_n}{\varepsilon}s$ makes all of these integrals solvable in closed form, allowing us to determine the ideal effective action.

Before doing these explicit calculations, we first look at which terms need to be retained in order to get the sought-after $1/N^p$ convergence. It is easy to show that the ideal effective action is a sum of terms of the form
\begin{equation}
\varepsilon^\alpha\,\delta^{2\beta}\Big(\partial^{\gamma_1} V(\bar q_n)\Big)\cdots \Big(\partial^{\gamma_k}V(\bar q_n)\Big)\, ,
\end{equation}
where $\alpha$, $\beta$, and $\gamma_1,\ldots,\gamma_k$ are nonnegative whole numbers constrained by simple dimensional analysis to satisfy \begin{equation}
\alpha+\beta=k+\frac{1}{2}\,\sum_{r=1}^k\gamma_r\, .
\end{equation}
Short time propagation satisfies the diffusion relation $\delta_{n}^2\propto \varepsilon$. Thus, $1/N^p$ convergence follows from  keeping all the terms satisfying $\alpha+\beta<p+1$. An equivalent, but practically more useful, form of the criterion determining the relevant terms at level $p$ is
\begin{equation}
k+\frac{1}{2}\,\sum_{r=1}^k\gamma_r<p+1\, .
\end{equation}

We now continue with the explicit evaluation of the ideal effective action, illustrating the general procedure on calculations to level $p=2$. The action is now
\begin{equation}
 S_n[\xi]=\varepsilon \left(\frac{1}{2}\,\frac{\delta_n^2}{\varepsilon^2}+V(\bar q_n)+\frac{\delta_{n,i}\delta_{n,j}}{24}\partial_{ij}^2V(\bar q_n)\right)+O(\varepsilon^3)\, .
\end{equation}
Note that $\delta_{n,i}$ is the $i$-th component of vector $\delta_n$ while $\partial^2_{ij}$ is shorthand for $\partial_i\partial_j$. The first two terms in the above expression correspond to the naive action, while the third term gives contributions of order $\varepsilon^2$. The remaining contribution at level $p=2$ comes from the expectation value
\begin{eqnarray}
\left\langle e^{ -\int ds\ U(x;\xi)}\right\rangle&=&1-\int_{-\frac{\varepsilon}{2}}^{\frac{\varepsilon}{2}}d s\ \left\langle U(x;\xi) \right\rangle+O(\varepsilon^3)=\nonumber\\
&=&1-\int_{-\frac{\varepsilon}{2}}^{\frac{\varepsilon}{2}} ds\ \frac{1}{2}\ \langle x_i x_j \rangle\ \partial_i \partial_j V(\xi) +O(\varepsilon^3)=\nonumber\\
&=&1-\frac{1}{2}\ \delta_{ij}\int_{-\frac{\varepsilon}{2}}^{\frac{\varepsilon}{2}} ds\ \Delta(s,s)_{i,j}\ \partial_{ij}^2 V(\xi)+O(\varepsilon^3)=\nonumber\\
&=&1-\frac{1}{2}\ \partial^2 V(\bar q_n)\int_{-\frac{\varepsilon}{2}}^{\frac{\varepsilon}{2}} ds\ \Delta(s,s)_{i,i}\ +O(\varepsilon^3)\, .\nonumber
\end{eqnarray}
As promised, the remaining integral is easily evaluated in closed form to yield $\varepsilon^2/6$. Appendix A lists all the related integrals needed for higher level calculations. The expectation value now equals
\begin{equation}
\left\langle e^{ -\int d s\ U(x;\xi)}\right\rangle = 1-\frac{\varepsilon^2}{12}\partial^2 V(\bar{q}_n)+O(\varepsilon^3) = e^{-\frac{\varepsilon^2}{12}\partial^2 V(\bar{q}_n)}+O(\varepsilon^3)\, .
\end{equation}
Finally, the $p=2$ level discretized effective action is simply
\begin{equation}
S_N^{(p=2)}=\sum_{n=0}^{N-1}\varepsilon\left[ \frac{1}{2}{\left(\frac{\delta_n}{\varepsilon}\right)}^2+V(\bar{q}_n)+
\frac{\varepsilon}{12}\,\partial^2 V(\bar{q}_n)+\frac{\delta_{n,i} \delta_{n,j}}{24}\,\partial_{ij}^2 V(\bar{q}_n)\right]\, .
\end{equation}
One similarly obtains the higher $p$ level effective actions. Appendix~B gives the explicit analytical expressions for many-particle discretized effective actions at level $p\le 5$. There are no obstacles in going to higher values of $p$. The derived expressions become algebraically more complex, and calculations are best done using a package for symbolic calculus such as Mathematica. Higher level effective actions can be found on our web site \cite{speedup}.

\section{Numerical results}
\label{sec:num}
In this section we present results of numerical PIMC simulations that confirm the analytically derived speedup in convergence of discretized path integrals. To do this, we have conducted a series of PIMC simulations of transition amplitudes for a two-dimensional system of two particles interacting through potential
\begin{equation}
V(\vec{r}_1,\vec{r}_2)=\frac{1}{2}(\vec{r}_1-\vec{r}_2)^2+\frac{g_1}{24}(\vec{r}_1-\vec{r}_2)^4+\frac{g_2}{2}(\vec{r}_1+\vec{r}_2)^2\, .
\label{eq:potential}
\end{equation}

\begin{figure}[!ht]
\centering
\includegraphics[width=0.8 \textwidth]{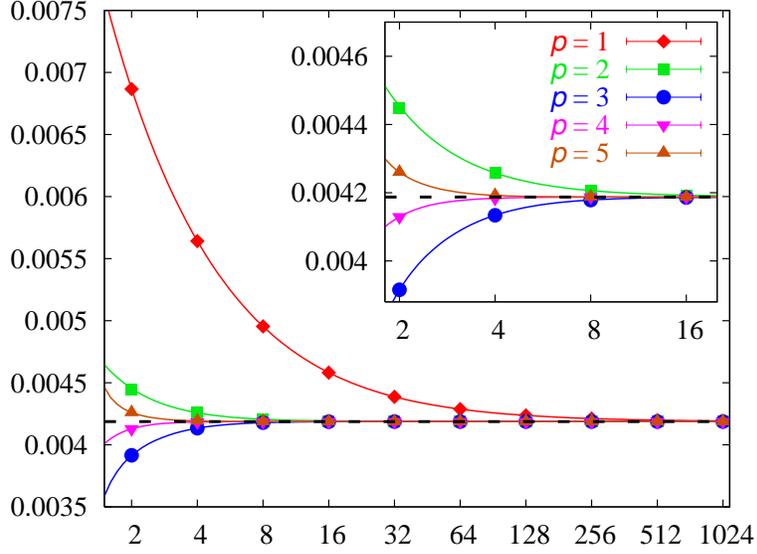}
\caption{Convergence of discretized amplitudes $A_N^{(p)}$ to the continuum as functions of $N$ for $p=1,2,3,4,5$ for a system of two particles in two dimensions moving in the quartic potential given in (\ref{eq:potential}) with coupling $g_1=10$, $g_2=0$, time of propagation $T=1$, and initial and final states $a=(0,0;0.2,0.5)$, $b=(1,1;0.3,0.6)$. The number of MC samples was $10^6$. The horizontal dashed line represents the continuum limit, solid lines correspond to the fitted functions (\ref{eq:fit}).}
\label{fig:1}
\end{figure}
\begin{figure}[!ht]
\centering
\includegraphics[width=0.8 \textwidth]{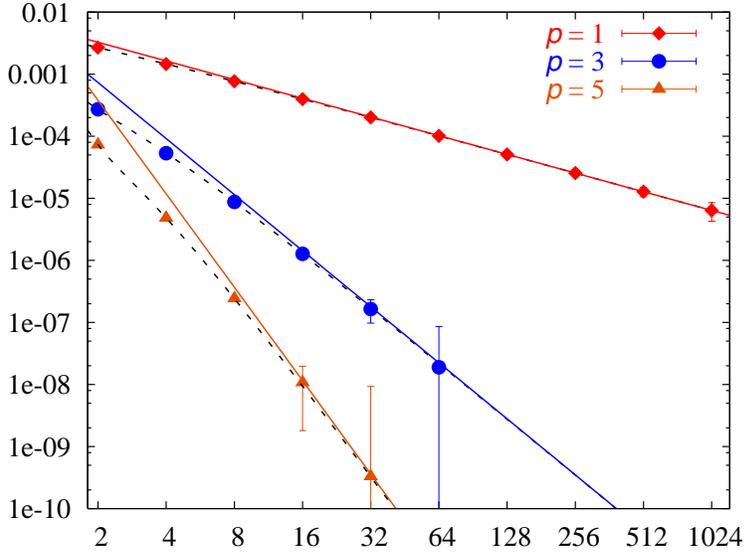}
\caption{Deviations from the continuum limit $|A_N^{(p)}-A|$ as functions of $N$ for $p=1,3,5$ for the system of two particles in two dimensions in the quartic potential given in (\ref{eq:potential}) with $g_1=10$, $g_2=0$, time of propagation $T=1$, initial and final states $a=(0,0;0.2,0.5)$, $b=(1,1;0.3,0.6)$. The number of MC samples was $10^6$ ($p=1$), $10^9$ ($p=3$), $10^{11}$ ($p=5$). Solid lines give the leading $1/N^p$ behavior, dashed lines correspond to the fitted functions (\ref{eq:fit}).}
\label{fig:2}
\end{figure}

Numerical simulations, based on our SPEEDUP \cite{speedup} PIMC code, have been performed for different values of couplings $g_1$ and $g_2$ and for variety of initial and final states. The associated continuum limit amplitudes $A^{(p)}$ have been estimated by fitting polynomials in $1/N$ to the discretized values $A_N^{(p)}$, according to the analytically derived relation (\ref{eq:impcon}):
\begin{equation}
A_N^{(p)}=A^{(p)}+\frac{B^{(p)}}{N^p}+\frac{C^{(p)}}{N^{p+1}}+\ldots\, .
\label{eq:fit}
\end{equation}
For all values of $p$ the fitted continuum values $A^{(p)}$ agree within the error bars. The obtained $1/N$ dependence gives explicit verification of the analytically derived increase in convergence. The typical case is illustrated in Figure \ref{fig:1}. The expected $1/N^p$ convergence may be most easily discerned from Figure \ref{fig:2}, a plot of the deviations of discretized amplitudes from the continuum limit. As can be seen, the increase of level $p$ leads to an ever faster approach to the continuum.

As a result of the newly presented method, the usual simulations (in which one calculates specific physical quantities such as the one in Figure \ref{fig:1}) proceed much faster than by using standard methods. On the other hand, Figure \ref{fig:2} is itself time consuming since it illustrates subdominant behavior. For this reason the figures contain only results obtained by effective actions to level $p=5$. Note, however, that the $p=5$ curve corresponds to a precision of four decimal places even for an extremely coarse discretization such as $N=2$.

In usual PIMC calculations one always chooses the number of MC samples so that the stochastic error of numerical results is of the order of deviations from the continuum limit. In Figure \ref{fig:2} the number of MC samples had to be much larger, in order for deviations from the continuum limit to be clearly visible.

\section{Energy spectra}
\label{sec:spec}

The improved convergence of path integral expressions for amplitudes leads directly to the same kind of improvement in the convergence of discretized partition functions, owing to the relation
\begin{equation}
Z_N(\beta)=\int d q A_N(q,q;\beta)\, ,
\end{equation}
where the inverse temperature $\beta$ plays the role of the time of propagation $T$. From the previous relation we directly obtain the path-integral presentation of the partition function:
\begin{equation}
Z_N(a,b;T)=\frac{1}{(2 \pi \varepsilon)^{\frac{M N d}{2}}}\int d q_1\cdots d q_{N}\ e^{-S_N}.\
\label{eq:zbase}
\end{equation}
The partition function is the central object for obtaining information about statistical systems. In addition, the partition function offers a straightforward way for extracting information about low lying energy levels of a system. This follows from evaluating the trace in the definition of the partition function in the energy eigen-basis
\begin{equation}
 Z(\beta)=\sum_{n=0}^{\infty} d_n e^{-\beta E_n}\, ,
\end{equation}
where $E_n$ and $d_n$ denote corresponding energy levels and degeneracies. A detailed description of the procedure for extracting energy levels from the partition function is given in \cite{stojiljkovicpla}. The free energy of the system, given by
\begin{equation}
 F(\beta)=-\frac{1}{\beta}\ln Z(\beta)\, ,
\end{equation}
tends to the ground state energy $E_0$ in the large $\beta$ limit. Similarly, we introduce auxiliary functions
\begin{equation}
 F^{(n)}(\beta)=-\frac{1}{\beta}\ln\frac{Z(\beta)-\sum_{i=0}^{n-1} d_i\ e^{-\beta E_i}}{d_n}\, ,
\end{equation}
which can be fitted for large $\beta$ to
\begin{equation}
f^{(n)}(\beta)=E_n-\frac{1}{\beta}\ln(1+a e^{-\beta b})\, .
\end{equation}
and which tend to the corresponding energy level $E_n$.

In this way, by studying the large $\beta$ behavior of the free energy and the corresponding auxiliary functions, one obtains the low lying energy spectrum of the model. However, in numerical simulations we are inevitably limited to a finite range of inverse temperatures. In addition, the above procedure for the construction of the auxiliary functions $f_n$ is recursive, i.e. in order to construct $f_n$ we need to know all the energy levels below $E_n$. This leads to the accumulation of errors as $n$ increases, practically limiting the number of energy levels that can be calculated. Note that the orders of magnitude increase in precision of the presented method reduces the magnitude of the accumulated error and thus allows us to extract a larger number of energy levels.

\begin{figure}[!ht]
\centering
\includegraphics[width=0.8 \textwidth]{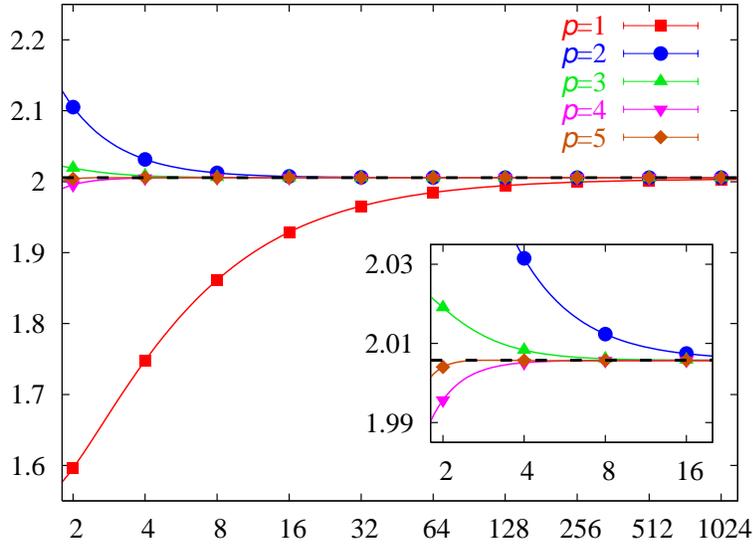}
\caption{Convergence of discretized free energies $F^{(p)}_N(\beta)$ to the continuum as functions of $N$ for $p=1,2,3,4,5$ for the system of two particles in two dimensions in potential (\ref{eq:potential}) for $g_1=1$, $g_2=1$, $\beta=1$. The number of MC samples was $10^{7}$. The horizontal dashed line represents the exact free energy.}
\label{fig:3}
\end{figure}
\begin{figure}[!ht]
\centering
\includegraphics[width=0.8 \textwidth]{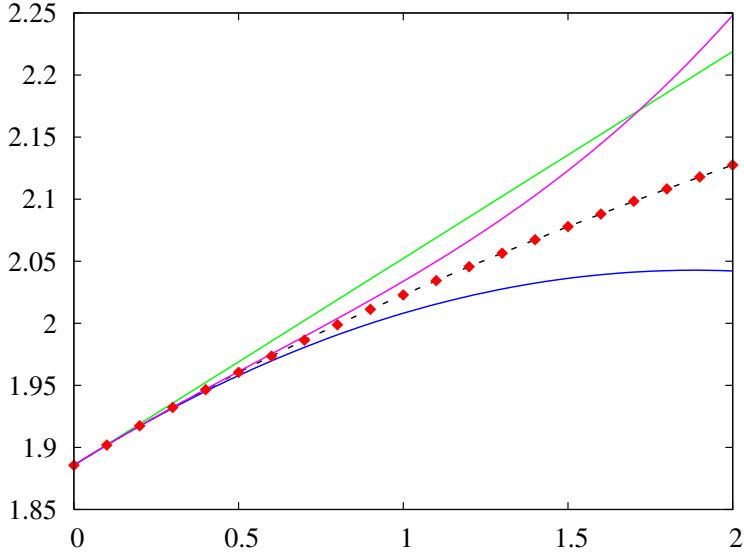}
\caption{Ground state energy versus coupling constant $g_1$ for two-dimensional two-particle system in the quartic potential with $g_2=1/9$. Numerical simulations were performed with $10^9$ MC samples, level $p=5$ effective action and $N=64$. Full lines give perturbative expansion results of the first order (green), second order (blue), and third order (pink). }
\label{fig:4}
\end{figure}

Figures \ref{fig:3} and \ref{fig:4} illustrate the calculation of low lying energy levels using the more efficient PIMC formalism presented in this paper on the case of a two-dimensional system of two distinguishable particles interacting through a quartic potential (\ref{eq:potential}). Figure \ref{fig:3} demonstrates that the improved convergence of free energies is the same as in the case of amplitudes. Table \ref{tab:table} gives the calculated energy levels for quartic coupling from $g_1=0$ (free theory) to $g_1=10$ (strongly interacting theory).

\begin{table}[!ht]
\begin{center}
\begin{tabular}{c|cccc}
\hline
$g_1$ & $E_0$ & $E_1$ & $E_2$ & $E_3$\\
\hline\hline
0.0&$ 1.8857(1)$&$2.3571(6)$&$2.83(1)$&$3.3(2) $\\
\hline
0.1&$ 1.9019(2)$&$2.374(2)$&$2.82(1)$&---\\
\hline
1.0 &$ 2.0228(2) $&$2.497(3)$&$2.94(3)$&---\\
\hline
10 &$ 2.6327(6)  $&$3.098(4)$&$3.57(3)$&---\\
\hline
\end{tabular}
\end{center}
\caption{Low lying energy levels for two-dimensional two-particle system interacting through potential (\ref{eq:potential}) with $g_2=1/9$. Calculations were done with $10^9$ MC samples, level $p=5$ effective action and $N=64$. The degeneracies of the calculated energy levels were found to be $d_0=1$, $d_1=2$, $d_2=3$, $d_3=6$.}
\label{tab:table}
\end{table}

Figure \ref{fig:4} presents a comparison of ground energy values calculated using the derived effective actions with those calculated by perturbative expansion. The graph gives the dependence of obtained values of ground energy on the coupling constant $g_1$. The agreement with perturbative results is excellent for small values of the coupling. Around $g_1\sim 1$ even the higher order perturbative results start deviating substantially from the exact result, while for larger couplings we are well in the non-perturbative regime in which such expansions are useless. Throughout, the calculations based on the use of the derived effective actions converge to the exact ground energy even for relatively rough discretizations. From Table \ref{tab:table} we see that this holds even in the case of strongly interacting theory.

\section{Energy estimators}
\label{sec:est}

One of the basic properties of physical systems is the internal energy given as the energy expectation value. Various ways of evaluating energy expectation values in PIMC simulations are based on different types of energy estimators -- expressions whose thermal expectation values are used to calculate the discretized internal energy. The practical choice of the best estimator usually depends on the model and on the specific values of physical parameters. It is determined through a comparison of the properties of different estimators (e.g. variance, convergence, etc.), both in terms of calculated numerical values and their functional dependencies. Details of the derivation and characteristics of several energy estimators are given in \cite{ceperley} and \cite{jankesauer}. The virial energy estimator turns out to be most advantageous due to its roughly constant Monte Carlo variance with the increase in the number of time slices. In this section we construct a hierarchy of virial energy estimators with improved convergence to the continuum limit as $1/N^p$, while keeping the feature of constant variance of the naive virial estimator.

First we briefly review the standard derivation of the virial estimator \cite{jankesauer,grujicpla}. Starting from the formula $U(\beta)=-\partial_{\beta} \log Z(\beta)$, the straightforward generalization to discretized expressions reads:
\begin{equation}
 U_N(\beta)=-\frac{\partial \log Z_N(\beta)}{\partial \beta}.
\end{equation}
The above partial derivative can be rewritten in terms of $\varepsilon$ as $\partial_{\beta}=1/N \; \partial_ \varepsilon$. In order to simplify the calculation of this derivative, we remove the
$\varepsilon$ dependence of the path integral measure and of the kinetic term of the discretized expression for $Z_N$ by simply rescaling $q\rightarrow q \sqrt{\varepsilon}$. Now the differentiation over $\varepsilon$ only affects the rescaled potential term in the exponent of the expression for $Z_N$. After the differentiation, we reinstall the original variables $q$, and obtain:
\begin{equation}
 U_N(\beta)=\frac{1}{Z_N}\left(\frac{1}{2 \pi \varepsilon}\right)^{\frac{N}{2}}\int dq_1 \ldots dq_N\,E_{virial}\,e^{-S_N}.
\label{eq:dis}
\end{equation}
Here, $E_{virial} $ is the standard virial energy estimator given as
\begin{equation}
 E_{virial}=\frac{1}{N} \sum_{n=0}^{N-1}\left(V(\bar q_n)+\frac{1}{2}\, \bar q_{n,i}\, \partial_i V(\bar q_n) \right).
\end{equation}
This estimator yields a typical $1/N$ convergence.

As shown in \cite{grujicpla}, improved convergence of expectation values follows from using the appropriate effective actions and energy estimators. As we have seen, virial estimators follow directly from the form of the discretized action. The level $p$ estimator is obtained by substituting $S_N^{(p)}$ for $S_N$ in this procedure. Writing the ideal effective action and its associated virial energy estimator as
\begin{eqnarray}
S^*_N &=& S_N+\sum_{p=2}^\infty \sigma^{(p)}\label{eq:s*} \\
E^*_{virial} &=& E_{virial}+\sum_{p=2}^\infty e^{(p)}\ ,
\end{eqnarray}
where $\sigma^{(p)}$ and $e^{(p)}$ are corresponding contributions proportional to $\varepsilon^p$. Each $\sigma^{(p)}$ is a sum over all the time-slices, i.e. $\sigma^{(p)}=\sum_{n=0}^{N-1}\sigma^{(p)}_n$. The same relation holds between $e^{(p)}$ and $e^{(p)}_n$. The outlined procedure now gives the following simple connection between ideal action and estimator:
\begin{equation}
e^{(p)}_n=\frac{1}{T}\,\left(p+\frac{1}{2}\,\bar q_{n,i}\,\partial_i\right)\sigma^{(p)}_n\ .
\end{equation}
\begin{figure}[!ht]
\centering
\includegraphics[width=0.85 \textwidth]{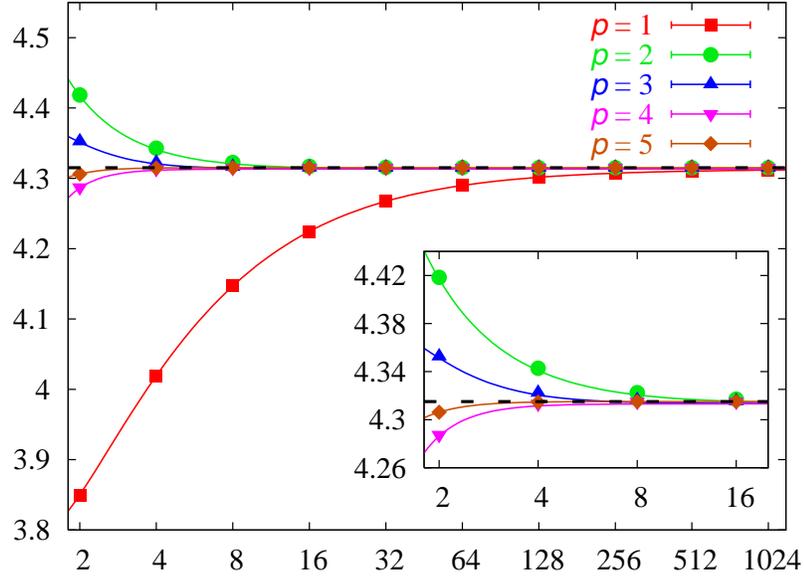}
\caption{Convergence of discretized energy expectation values $U^{(p)}_N(\beta)$ to the continuum as functions of $N$ for $p=1,2,3,4,5$ for the system of two particles in two dimensions in a potential (\ref{eq:potential}), with $g_1=1$, $g_2=1/9$, $\beta=1$. The number of MC samples was $10^{7}$. The horizontal dashed line represents the exact internal energy.}
\label{fig:5}
\end{figure}
\begin{figure}[!ht]
\centering
\includegraphics[width=0.85 \textwidth]{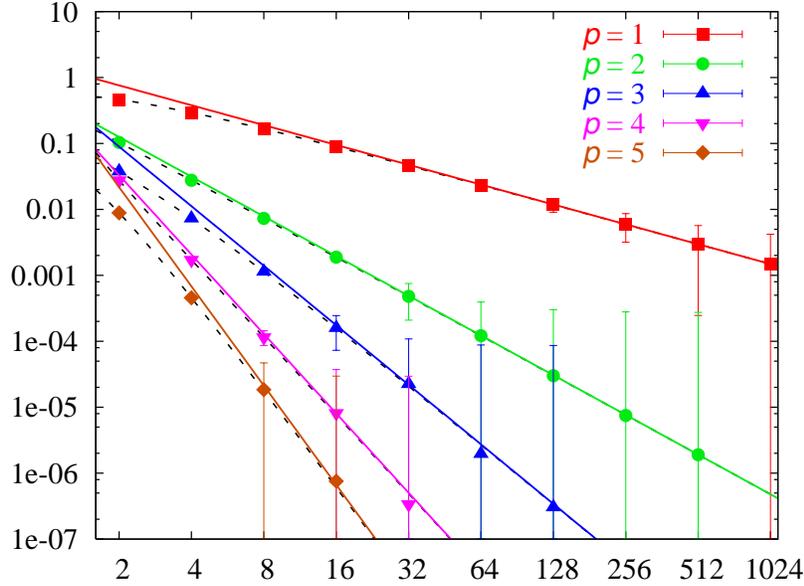}
\caption{Deviations from the continuum limit $|U_N^{(p)}(\beta)-U(\beta)|$ as functions of $N$ for $p=1,2,3,4,5$ for the system of two particles in two dimensions in quartic potential (\ref{eq:potential}), with $g_1=1$, $g_2=1/9$, $\beta=1$. The number of MC samples was $10^7$ ($p=1$), $10^9$ ($p=2$), $10^{10}$ ($p=3$), $10^{11}$ ($p=4,5$). Solid lines give the leading $1/N^p$ behavior, dashed lines correspond to the fitted functions (\ref{eq:fit}).}
\label{fig:6}
\end{figure}

The analytically derived improvement in convergence has been verified by a series of Monte Carlo simulations that have been performed on a two particle two dimensional system in a quartic potential (\ref{eq:potential}) for a variety of different values of coupling constant and inverse temperature. Typical results are presented in Figures \ref{fig:5} and \ref{fig:6}. Figure \ref{fig:5} shows convergence of discretized energy expectation values for different levels $p$. As level $p$ increases, we see that the continuum limit is approached faster, with ever smaller values of $N$. Numerical results conform precisely to the analytically derived $1/N^p$ increase in convergence as can be seen from the log-log plot of the deviations of discretized values from the continuum limit shown in Figure \ref{fig:6}.

\eject

\section{Conclusions}
\label{sec:con}

We have presented a derivation of discretized effective actions and energy estimators that lead to substantial, systematic speedup of numerical procedures for the calculation of path integrals of a generic many-particle non-relativistic theory. The derived speedup holds for all path integrals -- for transition amplitudes, partition functions, expectation values, as well as for calculations of energy levels. The obtained analytical results have been numerically verified through simulations of path integrals for  multi-particle model with quartic coupling. In the paper we present explicit expressions for the effective actions up to level $p=5$. The developed calculation scheme has been completed and is ready for applications to relevant problems in condensed matter physics. Further analytical work will focus on the generalization of the outlined scheme to more complex bosonic and fermionic systems, e.g. field theories.

\section*{Acknowledgements}

This work was supported in part by the Ministry of Science of the Republic of Serbia, under project No. OI141035, and the European Commission under EU Centre of Excellence grant CX-CMCS. The numerical results were obtained on the AEGIS e-Infrastructure, which is supported in part by FP6 projects EGEE-II and SEE-GRID-2.

\appendix
\onecolumn
\newpage

\section*{Appendix A: Integrals of propagators}

In this appendix we present the values of all integrals necessary for calculations of effective actions at levels $p\le 5$.
$$
\int_{-\frac{\varepsilon}{2}}^{\frac{\varepsilon}{2}} d s\ \Delta(s,s)^k\ s^{n-2k}=\left\{
\begin{array}{ll}
 0 & \qquad n \mbox{ odd}\\
2^{-1-n}\ \varepsilon^{1-k+n}\ B(1+k,\frac{1-2k+n}{2}) & \qquad n \mbox{ even ,}
\end{array}
\right.
$$
where $B$ is the Euler beta function. All the multiple integrals needed have the same boundaries of integration as the preceding case.

\begin{tabular}{ll}
$\int dsds'\ \Delta(s,s')=\frac{\varepsilon^3}{12} $& $\int dsds'\ \Delta(s,s')^3=\frac{\varepsilon^5}{560}$\\
$\int dsds'\ \Delta(s,s') \Delta(s',s')=\frac{\varepsilon^4}{60}$ & $\int dsds'\ \Delta(s,s') \Delta(s,s')=\frac{\varepsilon^4}{90}$\\
$\int dsds'\ s^2\ \Delta(s,s')=\frac{\varepsilon^5}{240}$ & $\int dsds'\ ss'\ \Delta(s,s')=\frac{\varepsilon^5}{720}$\\
$\int dsds'\ ss'^3\ \Delta(s,s')=\frac{\varepsilon^7}{6720}$& $\int dsds'\ s^2s'^2\ \Delta(s,s')=\frac{\varepsilon^7}{4032}$\\
$\int dsds'\ ss'\ \Delta(s,s')^2=\frac{\varepsilon^6}{5040}$&$\int dsds'\ s^4\ \Delta(s,s')=\frac{\varepsilon^7}{2240}$\\
$\int dsds'\ s'^2\ \Delta(s,s')^2=\frac{\varepsilon^6}{2520}$&$ \int dsds'\ s^2\ \Delta(s',s') \Delta(s,s')=\frac{\varepsilon^6}{1260}$\\
$\int dsds'\ s^2\ \Delta(s,s) \Delta(s,s')=\frac{\varepsilon^6}{1680}$& $\int dsds'\ ss'\ \Delta(s,s')\Delta(s',s')=\frac{\varepsilon^6}{5040}$\\
$\int dsds'\ \Delta(s,s)\Delta(s,s') \Delta(s',s')=\frac{17\varepsilon^5}{5040}$&$\int dsds'\ \Delta(s,s)\Delta(s,s')^2=\frac{\varepsilon^5}{420}$\\
$\int dsds'\ \Delta(s,s')\Delta(s,s)^2=\frac{\varepsilon^5}{280}$&$\int dsds'ds''\ \Delta(s,s'')\Delta(s',s'')=\frac{\varepsilon^5}{80}$\\
\end{tabular}

\section*{Appendix B: Effective actions for levels $p\le 5$}

The ideal effective action for a general non-relativistic multi-particle system in the mid-point prescription is given in (\ref{eq:s*}) as a sum of terms $ \sigma^{(p)} $ containing all contributions of order $\varepsilon^p$. As we have seen, $\sigma^{(p)}$ is the sum over all the time-slices, i.e. $\sigma^{(p)}=\sum_{n=0}^{N-1}\sigma^{(p)}_n$, where $\sigma^{(p)}_n$ is the contribution of time-slice $n$. In this appendix we give a list of the explicit expressions for $\sigma^{(p)}_n$ up to $p=5$ in shorthand notation in which $V=V(\bar{q}_n)$ and $\delta_i=\delta_{n,i}$. The expressions for higher levels can be found on our web site \cite{speedup}.

\begin{eqnarray}
\sigma^{(2)} & = &\frac{\varepsilon^2}{12}\partial^{2} V+\frac{\varepsilon \delta_i \delta_j}{24}\partial_{ij}^2 V\nonumber\\
\sigma^{(3)} & = &-\frac{{\varepsilon}^3}{24}\partial_i V\partial_i V+\frac{{\varepsilon}^3}{240}\partial^{4} V+
     \frac{\varepsilon^2 \delta_i \delta_j}{480}\partial_{ij}^2 \partial^{2} V+
     \frac{\varepsilon \delta_i \delta_j \delta_k \delta_l }{1920} \partial_{ijkl}^4 V \nonumber\\
\sigma^{(4)} & = &
\frac{{\varepsilon}^4}{6720}\partial^6 V-
\frac{{\varepsilon}^4}{120} \partial_iV \partial_i \partial^{2} V-
\frac{{\varepsilon}^4}{360} \partial_{ij}^2 V \partial_{ij}^2 V
-\frac{{\varepsilon}^3 \delta_i \delta_j }{480} \partial_k V \partial_{kij}^3 V+\nonumber\\
&+&
\frac{{\varepsilon}^3 \delta_i \delta_j}{13 440}\partial_{ij}^2 \partial^4 V
-\frac{{\varepsilon}^3 \delta_i \delta_j}{1440}\partial_{ik}^2 V \partial_{kj}^2 V
+\frac{\varepsilon^2 \delta_i \delta_j \delta_k \delta_l}{53760}\partial_{ijkl}^4\partial^{2} V+\nonumber\\
&+&
\frac{\varepsilon \delta_i \delta_j \delta_k \delta_l \delta_m \delta_n}{322 560}\partial_{ijklmn}^6 V\nonumber\\
\sigma^{(5)} & = &
\frac{\varepsilon^5}{241920}\partial^8 V -\frac{\varepsilon^5}{1680}\partial_{ij}^2V \partial_{ij}^2\partial^{2}V-
\frac{17 \varepsilon^5}{40320}\partial_i\partial^{2}V \partial_i\partial^{2}V-\nonumber\\
&-&
\frac{\varepsilon^5}{2240}\partial_iV\partial_{i}\partial^4 V
-\frac{\varepsilon^5}{6720}\partial_{ijk}^3 V \partial_{ijk}^3 V
+\frac{\varepsilon^5}{240}\partial_i V \partial_j V \partial_{ij}^2 V+\nonumber\\
&+&
\frac{\varepsilon^4 \delta_i\delta_j}{483 840}\partial_{ij}^2 \partial^6 V
- \frac{\varepsilon^4 \delta_i\delta_j}{6720}\partial_k V \partial_{ijk}^3 \partial^{2}V -
\frac{\varepsilon^4 \delta_i\delta_j}{10080}\partial_{ik}^2 V\partial_{kj}^2 \partial^{2}V-\nonumber\\
&-&
\frac{\varepsilon^4 \delta_i\delta_j}{10080}\partial_{kl}^2V\partial_{ijkl}^4 V
- \frac{\varepsilon^4 \delta_i\delta_j}{5040}\partial_k\partial^{2}V\partial_{ijk}^3 V-
\frac{\varepsilon^4 \delta_i\delta_j}{20160}\partial_{ikl}^3 V \partial_{klj}^3 V+\nonumber\\
&+&
\frac{\varepsilon^3 \delta_i \delta_j \delta_k \delta_l}{1935360} \partial_{ijkl}^4\partial^4 V
-\frac{\varepsilon^3 \delta_i \delta_j \delta_k \delta_l}{53760} \partial_m V \partial_{mijkl}^5 V-\nonumber\\
 &-&
\frac{\varepsilon^3 \delta_i \delta_j \delta_k \delta_l}{40320}\partial_{im}^2 V\partial_{mjkl}^4 V
-\frac{\varepsilon^3 \delta_i \delta_j \delta_k \delta_l}{32256}\partial_{ijm}^3 V\partial_{mkl}^3 V+\nonumber\\
&+&
\frac{\varepsilon^2 \delta_i \delta_j \delta_k \delta_l \delta_m \delta_n }{11612160}\partial_{ijklmn}^6 \partial^{2}V
+\frac{\varepsilon \delta_i \delta_j \delta_k \delta_l \delta_m \delta_n \delta_p \delta_q}{92897280}\partial_{ijklmnpq}^8 V\nonumber
\end{eqnarray}

\end{document}